\documentclass[12pt]{article}
\newcommand{\be}{\begin{equation}}
\newcommand{\ee}{\end{equation}}
\newcommand{\bea}{\begin{eqnarray}}
\newcommand{\eea}{\end{eqnarray}}

\begin{document}

\renewcommand{\thefootnote}{\fnsymbol{footnote}}
\renewcommand{\baselinestretch}{1.3}
\medskip

\begin{center}
{\large {\bf Can Black Holes Decay to Naked Singularities?}
\footnote{This essay received an "honorable mention" in the 2001 Essay
    Competition of the Gravity Research Foundation -- Ed.
}}
 \\ \medskip  {}
\bigskip
\bigskip

{\bf 
Saurya Das$^\sharp$, 
J. Gegenberg$^\dagger$, and  V. Husain$^\dagger$
\\}
 
{\sl
$^\sharp$ Dept. of Physics,
University of Winnipeg\\
Winnipeg, Manitoba, Canada R3B 2E9\\
 }
{\sl
$^\dagger$ Dept. of Mathematics and Statistics,
University of New Brunswick\\
Fredericton, New Brunswick, Canada  E3B 5A3
\footnote{emails: saurya@theory.uwinnipeg.ca, lenin@math.unb.ca, husain@math.unb.ca}\\
}
\end{center}

\bigskip
\bigskip
\centerline{\bf Abstract}
\smallskip 
 
\baselineskip = 2em 
\noindent We investigate thermodynamic properties of two types of asymptotically 
anti-de Sitter spacetimes: black holes and singular scalar field 
spacetimes. We describe the possibility that thermodynamic phase transitions 
can transform one spacetime into another, suggesting that black holes 
can  radiate to naked singularities.

\vfill
\eject

\baselineskip=2em

The discovery of the connection between semi-classical 
properties of black holes and thermodynamics has naturally 
led to the application of statistical physics ideas to black hole 
physics. The main questions concern the origin of black hole 
microstates and how these might arise from a quantum gravity theory, 
the final state of Hawking radiation, and the possibility 
of thermal phase transitions between black holes and other 
spacetimes. 
 
In this essay we discuss the last of these issues in the 
context of asymptotically anti-deSitter (AAdS) spacetimes. 
This class of spacetimes is of current interest  in the context 
of the AdS/CFT conjecture in string theory. 

A class of phase transitions for these spacetimes was studied by 
Hawking and Page \cite{hp}. They showed that the canonical ensemble 
exists, and that at a given temperature above a certain 
minimum, there are three possible states: a low mass black hole, a high 
mass black hole and thermal radiation in AdS space. The first of these 
states has negative specific heat and can decay to the other 
configurations, which are stable but not equally probable.

These thermodynamic results can be extended to arbitrary spacetime 
dimensions, and related to the confinement/deconfinement phase transition 
in a dual conformal Yang-Mills theory \cite{witten}. Similar phase transitions 
also occur between constant curvature black holes, whose horizon topologies 
are other than spherical, and the AdS soliton spacetime \cite{surya}.   

In all of these cases the possibility of a phase transition is indicated 
by a simple calculation: the Euclidean action difference $\Delta I$ of  
two spacetimes, which gives the probability ratio of the two 
configurations. If  $\Delta I$ changes sign as the 
metric parameters are varied, it is an indication that one phase 
becomes more probable than the other at a fixed temperature (determined 
by appropriate matching of Euclidean time periodicities).

We consider the Einstein-scalar field system $(g_{ab}, \phi)$, with minimal 
scalar field coupling, and show that there is a phase transition between black 
holes and naked singularities using arguments analogous to those 
used in Ref. \cite{hp}. We focus on  three dimensional spacetimes for which 
the details are particularly simple. 

The two Euclidean spacetimes of interest for this study are the 
BTZ black hole 
\be 
ds^2 = \left( (r/l)^2 - M\right) dt^2 + \left( (r/l)^2 -M\right)^{-1} dr^2 
        + r^2 d\theta^2,   
\ee
and the scalar field spacetime \cite{us}
\bea
ds^2 &=&  (x-b/2)^{(1+a)/2}(x+b/2)^{(1-a)/2} dt^2 +
 {l^2\over 4(x^2-b^2/4)} dx^2  \nonumber \\
       &+& \ell^2~(x-b/2)^{(1-a)/2}(x+b/2)^{(1+a)/2} d\theta^2, 
\label{newsol}
\eea
with 
\be
\phi(x) = \frac{1}{2}{\sqrt{{\frac{1-a^2}{2}}}}\ln\left(\frac{x-b/2}{x+b/2}, 
\right)
\label{scfield}
\ee
where $\ell$ is cosmological constant length scale. The latter solution
with $a=1$ 
is a BTZ black hole with $M_{BH} = b$. This makes the two metrics particularly 
easy to compare. 

To calculate the action difference $\Delta I$ for these two solutions one 
integrates out to a fixed value of the `radial' coordinate  $x=X$, ensuring that 
the Euclidean time periodicities for the $t$ integrals are fixed so that 
the local temperatures are equal there: $\beta_{SC}\sqrt{-g^{SC}_{00}} = 
\beta_{BH}\sqrt{-g^{BH}_{00}}$. The limit $X\rightarrow\infty$ is 
taken at the end. The scalar matter terms must of course be included in the $\Delta I$ 
calculation.  
 
At fixed temperature the action difference is 
\be
\Delta I := I_{SC} - I_{BH} = {\pi^2 l^2 \over 2r_+}\left((r_+/l)^2 -\sqrt{2}k \right),
\label{diff} 
\ee
where $r_+:={\sqrt{M}}l$ is the BTZ black hole radius, and $\sqrt{2}k = b\sqrt{1-a^2}$ is 
the scalar field strength (\ref{scfield}).  This is 
finite even though there is a naked singularity at the origin for the scalar 
field solution! $\Delta I$ changes sign depending on the size of the black 
hole $r_+$ in comparison with the scalar field strength $k$. This indicates a 
phase transition: {\it sufficiently small black holes phase transform semi-classically 
to naked singularities!}

This calculation suggests an outline (at the semi-classical level) of the 
entire process of gravitational collapse and its subsequent evolution. 
 A scalar field collapsing in an AdS spacetime 
will form a black hole if appropriate initial conditions are satisfied 
\cite{collapse}. The black hole will then radiate away its mass via 
Hawking radiation. When the mass is sufficiently small, of the order 
$k$, a phase change will occur to the  scalar field 
configuration characterized by $k$. This is a static geometry with 
a naked singularity. 

Several interesting questions can be asked at this stage. If the
(AdS)$\ \leftrightarrow\ $(Schwarzschild-AdS) transition reflects
the  (confinement)$\ \leftrightarrow\ $(deconfinement) phase change
in the boundary Yang-Mills theory, what is the correspondence for the
(black hole) $ \leftrightarrow\ $(naked singularity) transition
described above? Are there certain exotic states in the boundary
conformal field theory which are dual to naked singularities in the
bulk? Is a naked singularity indeed the final state of gravitational
collapse in this model as suggested by our semiclassical argument?
While some of these issues can be investigated in the currently
available framework, the resolution of the others, such as those
related to the end state of collapse would require one to go beyond
the semi-classical regime to a full quantum theory of gravity.

\end{document}